\documentstyle[12pt]{article}
\def\appendix{\par
 \setcounter{section}{0}
 \setcounter{subsection}{0}
 \def\thesection{\Alph{section}}
 \def\theequation{\thesection.\arabic{equation}}}
 \def\thebibliography#1{\subsection*{References}\list
 {[\arabic{enumi}]}{\settowidth\labelwidth{[#1]}
 \leftmargin\labelwidth
 \advance\leftmargin\labelsep
 \usecounter{enumi}}
 \def\newblock{\hskip .11em plus .33em minus .07em}
 \sloppy\clubpenalty4000\widowpenalty4000
 \sfcode`\.=1000\relax}

\def\a{\alpha}
\def\b{\beta}
\def\c{\chi}
\def\d{\delta}

\def\vf{\varphi}
\def\g{\gamma}

\def\j{\psi}

\def\l{\lambda}
\def\m{\mu}

\def\p{\pi}
\def\q{\theta}

\def\x{\xi}
\def\z{\zeta}

\def\cm{{\cal M}}

\def\cu{{\cal U}}
\def\cv{{\cal V}}

\def\inbar{\vrule height1.5ex width.4pt depth0pt}
\def\rlx{\relax\leavevmode}
\def\I{\leavevmode\hbox{\small1\kern-3.8pt\normalsize1}}
\def\openone{\leavevmode\hbox{\small1\kern-3.3pt\normalsize1}}
\def\Ione{\rlx{\rm 1\kern-2.7pt l}}
\font\cmss=cmss10
\font\cmsss=cmss10 at 7pt
\def\ZZ{\rlx\leavevmode
             \ifmmode\mathchoice
                    {\hbox{\cmss Z\kern-.4em Z}}
                    {\hbox{\cmss Z\kern-.4em Z}}
                    {\lower.9pt\hbox{\cmsss Z\kern-.36em Z}}
                    {\lower1.2pt\hbox{\cmsss Z\kern-.36em Z}}
               \else{\cmss Z\kern-.4em Z}\fi}
\def\Ik{\rlx{\rm I\kern-.18em k}}  
\def\IC{\rlx\leavevmode
             \ifmmode\mathchoice
                    {\hbox{\kern.33em\inbar\kern-.3em{\rm C}}}
                    {\hbox{\kern.33em\inbar\kern-.3em{\rm C}}}
                    {\hbox{\kern.28em\sinbar\kern-.25em{\rm C}}}
                    {\hbox{\kern.25em\ssinbar\kern-.22em{\rm C}}}
             \else{\hbox{\kern.3em\inbar\kern-.3em{\rm C}}}\fi}
\def\IP{\rlx{\rm I\kern-.18em P}}
\def\IR{\rlx{\rm I\kern-.18em R}}
\def\IN{\rlx{\rm I\kern-.20em N}}

\def\llsymbol#1{\@llsymbol{\@nameuse{c@#1}}}
\def\@llsymbol#1{\ifcase#1\or {}\or {'}\or {''}\or {'''}\or
   {''''}\or {'''''}\or  \else\@ctrerr\fi\relax}

\newcounter{contador}
\newcommand{\letra}{
   \stepcounter{equation}
   \setcounter{contador}{\value{equation}}
   \setcounter{equation}{0}
   \renewcommand{\theequation}{\thecontador.\alph{equation}}}
\newcommand{\antiletra}{
   \renewcommand{\theequation}{\arabic{equation}}
   \setcounter{equation}{\value{contador}}}

\newcommand{\ol}\overline
\newcommand{\ti}\tilde
\newcommand{\wt}\widetilde
\newcommand{\wh}\widehat
\newcommand{\bv}\breve
\newcommand{\dg}\dagger
\newcommand{\pari}{\stackrel{{P}}\longrightarrow}

\newcommand{\Ddd}{$D$$=$$1$$+$$2$}

\newcommand{\aand}{\;\;\;\mbox{and}\;\;\;}

\newcommand{\be}{\begin{equation}}
\newcommand{\ee}{\end{equation}}
\newcommand{\bl}{\begin{eqnarray}&}
\newcommand{\el}{&\end{eqnarray}}
\newcommand{\bq}{\begin{eqnarray}}
\newcommand{\eq}{\end{eqnarray}}

\newcommand{\ts}{\textstyle}

\newcommand{\sx}{\sigma_x}
\newcommand{\sy}{\sigma_y}
\newcommand{\sz}{\sigma_z}

\newcommand{\gm}{{\gamma}^m}

\newcommand{\ov}{\overline}
\newcommand{\pa}{\partial}

\newcommand{\ra}{\rangle}
\newcommand{\la}{\langle}

\def\sl#1{\rlap{\hbox{$\mskip 1 mu /$}}#1}      
\def\Sl#1{\rlap{\hbox{$\mskip 3 mu /$}}#1}      
\def\SL#1{\rlap{\hbox{$\mskip 4.5 mu /$}}#1}    

\begin{document}
\title{\Large \bf Electron-pair condensation in 
parity-preserving QED$_{3}${\thanks{Talk given at {\it Quantum 
Systems: 
New Trends and Methods 96 - QS96 - Minsk - Belarus.}}} }
\author{
{\it M. A. De Andrade}$^\dg$ ,~{\it O. M. 
Del Cima}{\thanks{ {\it Pontif\'\i cia Universidade 
Cat\'olica do Rio 
de Janeiro (PUC-RIO), Departamento de F\'\i sica, Rua 
Marqu\^es de S\~ao  
Vicente, 225 - G\'avea - 22453-900 - Rio de Janeiro - 
RJ - Brazil}. 
M.A.D.A. e-mail:marco@fis.puc-rio.br. O.M.D.C. 
e-mail:oswaldo@fis.puc-rio.br.}}  {\thanks{ {\it Centro 
Brasileiro 
de Pesquisas F\'\i sicas (CBPF), Departamento 
de Teoria de Campos e Part\'\i culas (DCP), Rua Dr. 
Xavier Sigaud, 150 
- Urca - 22290-180 - Rio de Janeiro - RJ - Brazil}. 
J.A.H.N. e-mail:
helayel@cbpfsu1.cat.cbpf.br.}} {\thanks{ Address after 
September 1, 
1997: {\it Institut f\"ur Theoretische Physik, 
Technische Universit\"at 
Wien, Wiedner Hauptstra{\ss}e 8-10 - A-1040 - Vienna - 
Austria}.}} 
~and~{\it J. A. Helay\"el-Neto}$^\ddagger$    } 
\date{}
\maketitle
\begin{abstract}
In this paper, we present a parity-preserving 
QED$_{3}$ with spontaneous
breaking of a local $U(1)$-symmetry. The breaking 
is accomplished by a
potential of the $\vf^6$-type. It is shown that a 
net attractive interaction
appears in the M{\o}ller scattering ($s$ and $p$-wave 
scattering
 between two 
electrons) as mediated by the gauge field and 
a Higgs scalar. This might
favour a pair-condensation mechanism.
\end{abstract}

\section{Introduction}

Over the past years, the study of 3-dimensional 
field theories {\cite{djt}} has
been
well-supported in view of the possibilities they 
open up for the setting of a
gauge-field-theoretical foundation in the description 
of Condensed Matter
phenomena, such as High-$T_{c}$ Superconductivity 
{\cite{hightc}} and Quantum
Hall Effect {\cite{hall}}. Abelian models such as 
QED$_{3}$ and
${\tau}_{3}$QED$_{3}$ {\cite{domavro,qedtau3}}
are some of the theoretical approaches proposed 
to describe more deeply some
features of
high-$T_{c}$ materials.

The theory of superconductivity by Bardeen, Cooper 
and Schrieffer (BCS model)
{\cite{bcs}} succeeds in providing a microscopical 
description for
superconducting materials: indeed, many predictions 
of the BCS model have been
confirmed experimentally. An elegant mathematical 
formulation was given to it
by Bogoliubov {\cite{bogo}}. The characteristic 
feature of the BCS theory is
that it produces an energy gap between the ground 
state and the excited states
of a superconductor. The gap is due to the fact 
that the attractive
phonon-mediated interaction between electrons
 produces correlated pairs of such
particles (Cooper pairs) {\cite{cooper}}, with 
opposite momenta and spin; a
finite amount of energy is required to break 
this correlation.

In a well-known paper by Nambu and Jona-Lasinio 
{\cite{dsb1}}, it was proposed
that the nucleon mass might arise from a dynamical 
mechanism, similar to the
appearance of the energy gap in the BCS model. 
They proposed that elementary
excitations in a superconductor could be described 
by means of a coherent
mixture of electrons and holes. The framework they 
set up for dynamical mass
generation was motivated by the observation of an 
analogy between the
properties of Dirac particles and the quasi-particle 
excitations that appear in
a superconductor.

The main purpose of this paper is to show that 
electrons scattered in {\Ddd}
can experience a mutual net attractive interaction, 
not depending on their spin states.
This attractive scattering potential comes from 
processes in which the
electrons are correlated in momentum space with 
opposite spin polarisations
($s$-wave state). Also, in the case of equal spin 
polarisations ($p$-wave state), a net attraction may appear,
as due to the Higgs interaction, if some special 
conditions are set up on the parameters.
The latter possibility should be investigated for the cases in 
which very high external magnetic fields are 
applied, since it is suspected that the resistance 
of the superconducting state in the 
presence of high magnetic fields, in the re-entrant 
superconductivity effect,
could be explained by $p$-wave states, 
$p$-electron pairing {\cite{high-b}}. The 
intermediate bosons are a 
massive vector meson and a
Higgs scalar, both resulting from the breaking 
of a local $U(1)$-symmetry. The
breakingdown is accomplished by a sixth-power 
potential. We analyse the
conditions on the parameters in order to avoid 
metastable vacuum states. The
method used here to compute the scattering potentials 
is based on the ideas
reported in a series of papers by Sucher {\it et al.} 
{\cite{sucher}}. The
behaviour of the scattering interactions mediated by 
the massive vector meson
and the Higgs scalar are presented for electrons 
scattered in $s$ and $p$-wave processes. The interesting 
feature of $s$ 
and $p$ scatterings, since net 
attractive potentials are generated, 
motivated the study of Bethe-Salpeter 
equation {\cite{bethesalp,bethesalp3}} associated to the
model proposed here {\cite{bsqed3ssb}}, in order to verify, if 
whether or not there are $s$ and $p$-wave bound states. 
The issue of 
confinement in QED$_{3}$ 
{\cite{maris}} is also alluded
to. The behaviour at the quantum level of the model 
proposed in this letter, 
in the symmetric and broken regimes, is analised in 
ref.{\cite{arqed3}} by 
using the algebraic renormalisation method, which is 
independent of any 
kind of regularisation scheme {\cite{algren}}.

The outline of our paper is as follows. In Section 2, the 
parity-preserving Abelian model is presented as the 
spontaneous breaking of the $U(1)$-symmetry is discussed 
in the $R_\x$-gauge. Next, in Section 3, the calculation of 
the tree-level scattering amplitude is presented and the
net attractive potential is worked out. Finally, in Section 4, 
we cast a few remarks on our results. One Appendix follows, 
where a few comments on Dirac fermions in {\Ddd} are pointed
out.

\section{Parity-preserving QED$_{3}$ coupled to scalar matter}

The action for the parity-preserving QED$_{3}$
\footnote{The metric adopted
throughout this work is $\eta_{mn}=(+,-,-)$; $m$,
$n$=(0,1,2). Note that slashed objects mean contraction 
with $\g$-matrices. The
latter are taken as $\g^m$$=$$(\sx,i\sy,-i\sz)$.} 
with spontaneous symmetry
breaking of a local $U(1)$-symmetry is given by :
\bq
S_{{\rm QED}}\!\!\!\!&=&\!\!\!\!\int{d^3 x}      
\left\{ -{1\over4}
F^{mn}F_{mn}
+ i {\ov\j _+} {\SL{D}} {\j}_+ + i
{\ov\j _-} {\SL{D}} {\j}_- - y (\ov\j_+\j_+ - 
\ov\j_-\j_-)\vf^*\vf \;+\right.
\nonumber\\
&&\left.
 +\;D^m\vf^* D_m\vf - V(\vf^*\vf){}^{}_{}\right\} 
\;\;\;\;\;\;\;,
\label{action1}
\eq
with the potential $V(\vf^*\vf)$ taken as
\be
V(\vf^*\vf)=\m^2\vf^*\vf + {\z\over2}(\vf^*\vf)^2 
+
{\l\over3}(\vf^*\vf)^3 \;\;\;\;\;\;, 
\label{potential}
\ee
where the mass dimensions of the parameters 
$\m$, $\z$, $\l$ and $y$ are
respectively ${1}$, ${1}$, ${0}$ and ${0}$.

The covariant derivatives are defined as follows :
\be
{\SL{D}}\j_{\pm}\equiv(\sl{\pa} + iqg \Sl{A})\j_{\pm} 
\aand
D_{m}\vf\equiv(\pa_{m} + iQ g A_{m})\vf \;\;\;, 
\label{covder}
\ee
where $g$ is a coupling constant with dimension of 
(mass)$^{1\over2}$ and, $q$
and $Q$ are the $U(1)$-charges of the fermions and 
scalar, respectively. In the
action (\ref{action1}), $F_{mn}$ is the usual field
strength for $A_m$, $\j_+$ and $\j_-$ are two kinds 
of fermions (the $\pm$
subscripts refer to their spin sign {\cite{binegar}}, 
see 
also the Appendix) 
and $\vf$ is a complex
scalar. The $U(1)$-symmetry gauged by $A_m$ is 
interpreted as the
electromagnetic one, so that $A_m$ is meant 
to describe 
the photon. It is noteworthy to remark that 
terms of the form
$\j^\a_{\pm}\j_{\pm\a}\vf\vf$ and 
$\j^\a_{\pm}\j_{\pm\a}\vf^*\vf^*$ are
not adjoined to the interaction Lagrangian because 
Lorentz
invariance would require the fermion to be Majorana
\footnote{For Dirac 
fermions ($\j$) one has 
${\ov\j}$$\equiv$${\ov\j}^{\a}$$=$$-C^{\a\b}\j^c_{\b}$, 
since 
for Majorana fermions ($\q$) $\q^c$$=$$\q$, then it follows 
that ${\ov\q}$$\equiv$${\ov\q}^{\a}$$=$$-C^{\a\b}\q_{\b}$ . 
Therefore, 
for Majorana fermions 
${\ov\q}\q$$=$$\q^{\a}\q_{\a}$ .}. 
However,
if such were the case these terms would explicitly break
the $U(1)$-invariance, unless there would be more than a 
flavour of scalars. So, since we are dealing with 
Dirac fermions and 
just a complex scalar, the term $\ov\j_{\pm}\j_{\pm}
\vf^*\vf$ is indeed 
the only one that
couples fermions to scalars in a way compatible with 
Lorentz and gauge invariance while respecting 
renormalisability.

The QED$_{3}$-action\footnote{For more details about 
QED$_{3}$ and
${\tau}_{3}$QED$_{3}$, as well as their applications 
and some peculiarities of
parity and time-reversal in {\Ddd}, see refs.
{\cite{djt,domavro,qedtau3}}.}
(\ref{action1}) is invariant under the discrete 
symmetry, $P$, whose action is
fixed below :
\letra
\bq
x_m &\pari& x_m^P=(x_0,-x_1,x_2)\;\;\;, \label{xp}\\
\j_{\pm} &\pari& \j_{\pm}^P=-i\g^1\j_{\mp}\;\;,\;\;\;
\ov\j_{\pm} ~\pari ~\ov\j_{\pm}^P=i\ov\j_{\mp}\g^1\;\;\;, 
\label{psip}\\
A_m &\pari& A_m^P=(A_0,-A_1,A_2)\;\;\;, \label{vfp}\\
\vf &\pari& \vf^P=\vf\;\;\;. \label{sp}
\eq
\antiletra
Since we are looking for a model that preserves the 
parity and time-reversal in
{\Ddd}, it should be noticed that the transformation 
(\ref{vfp}) has been
imposed in such a way that the interactions respect 
both invariances.

The sixth-power potential, $V$, is the responsible 
for breaking the
electromagnetic $U(1)$-symmetry. It is the most 
general renormalisable
potential in $3D$.

Analysing the potential (\ref{potential}), and 
imposing that it is bounded from
below and yields only stable vacua (metastability is 
ruled out), the following
conditions on the parameters $\m$, $\z$, $\l$ must be set :
\be
\l>0 \;\;,\;\;\; \z<0 \aand \m^2 \leq {3\over 16} 
{\z^2\over \l} \;\;\;.
\label{cond}
\ee
We denote ${\langle}\vf{\rangle}$$=$$v$ and the vacuum 
expectation value for
the $\vf^*\vf$-product, $v^2$, is chosen as
\be
{\langle}\vf^*\vf{\rangle}=v^2=-{\z \over 2\l}+ 
\left[ \biggl({\z \over
2\l}\biggr)^2 - {\m^2\over \l} \right]^{1\over 2} 
\;\;\;, \label{vac}
\ee
the condition for minimum being read as
\be
\m^2+{\z}v^2+{\l}v^4=0 \;\;\;. \label{mincond}
\ee

The complex scalar, $\vf$, is parametrised by
\be
\vf=v+H+i\q\;\;\;, \label{para}
\ee
where $\q$ is the would-be Goldstone boson and $H$ 
is the Higgs scalar, both
with vanishing vacuum expectation values. It should be 
noticed that the 
parametrisation given by eq.(\ref{para}) was chosen in 
order to avoid 
non-renormalisable interactions {\cite{priv1}}. 

By replacing the parametrisation (\ref{para}) for 
the complex scalar, $\vf$,
into the action (\ref{action1}), the following free 
action comes out:
\bq
{\hat S}_{{\rm QED}}^{\rm free}\!\!\!\!&=&\!\!\!\!
\int{d^3 x}\left\{-{1\over4}
F^{mn}F_{mn} + {1\over2} M^2_A A^m A_m
+  {\ov\j _+} (i{\sl{\pa} - m}) {\j}_+ +
{\ov\j _-} (i{\sl{\pa} + m}) {\j}_- \;+ \right. 
\nonumber\\
&&\left.
+ \;\pa^m H \pa_m H - M^2_H H^2+
{\pa^m}\q {\pa_m}\q + 2vQg{A^m} {\pa_m}\q \right\}
\;\;\;\;\;\;\;,\label{action2}
\eq
where the parameters $M^2_A$, $m$ and $M^2_H$ are 
given by
\be
M^2_A=2v^2Q^2g^2 \;\;,\;\;\;m=yv^2 \aand 
M^2_H=2v^2(\z+2 \l v^2) \;\;\;.\label{masses}
\ee
The conditions (\ref{cond}) and (\ref{mincond}) 
imply the following lower-bound
(see eq.(\ref{masses})) for the Higgs mass :
\be
M^2_H \geq {3\over 4} {\z^2\over \l} \;\;\;.
\label{lowbound}
\ee
Therefore, a {\it massless} Higgs is out of the model 
we consider here. A
massless Higgs would be present in the spectrum if
 $\m^2$$>$${3\over
16}{\z^2\over \l}$. But, in such a situation, 
the minima realising the
spontaneous symmetry breaking would not be absolute
ones, corresponding
therefore to metastable ground states, that we avoid 
here. One-particle states
would decay with a short decay-rate if compared to 
an absolute minimum ground
state.

In order to preserve the manifest renormalisability 
of the model, the 't Hooft
gauge {\cite{thooftg}} is adopted :
\bq
{\hat S}_{R_{\x}}^{\rm gf}\!\!\!\!&=&\!\!\!\!
\int{d^3 x}\left\{
-{1\over 2\x} \biggl( \pa^m A_m -{\sqrt{2}}\x 
M_A \q \biggl)^2 \right\}\;\;\;,
\label{gf}
\eq
where $\x$ is a dimensionless gauge parameter.

By replacing the parametrisation (\ref{para}) 
into the action (\ref{action1}),
and adding up the 't Hooft gauge (\ref{gf}), it 
can be directly found the
following complete parity-preserving action :
\bq
{S}_{{\rm QED}}^{\rm SSB}\!\!\!\!&=&\!\!\!\!
\int{d^3 x}\left\{-{1\over4}
F^{mn}F_{mn} + {1\over2} M^2_A A^m A_m
+  {\ov\j _+} (i{\sl{\pa} - m}) {\j}_+ +
{\ov\j _-} (i{\sl{\pa} + m}) {\j}_- \;+ \right. 
\nonumber\\
&&\left.
+ \;\pa^m H \pa_m H - M^2_H H^2+
{\pa^m}\q {\pa_m}\q - M^2_\q \q^2 -{1\over 2\x}
(\pa^m A_m)^2 \;+ \right. \nonumber \\
&&\left.
- \;qg {\ov\j _+}{\Sl{A}}{\j}_+ - qg {\ov\j _-}
{\Sl{A}}{\j}_-
-  y (\ov\j_+\j_+ - \ov\j_-\j_-) (2vH + H^2 +\q^2) \;+  
\right. \nonumber \\
&&\left. 
+ \;Q^2g^2 A^m A_m (2vH + H^2 +\q^2) + 
2QgA^m(H{\pa_m}\q - \q{\pa_m}H)
 \;+ \right. \nonumber \\
&&\left. -\; c_3 H^3 - c_4 H^4 - c_5 H^5 - c_6 
H^6 - c_7\q^4 - c_8\q^6 - c_9H\q^2 - c_{10}H^2\q^2 
\;+ \right. \nonumber \\
&&\left.
-\;c_{11}H^3\q^2 - c_{12}H^4\q^2 - c_{13}H\q^4 -  
c_{14}H^2\q^4 \right\}
\;\;\;\;\;\;\;,\label{action3}
\eq
where the constants $M^2_\q$, $c_3$, $c_4$, $c_5$, 
$c_6$, $c_7$, $c_8$, 
$c_9$, $c_{10}$, $c_{11}$, $c_{12}$, $c_{13}$ and 
$c_{14}$ are defined by
\bq
&&M^2_\q=\x M^2_A \;\;,\;\;\; c_3=2v(\z+{10\over3}\l 
v^2)\;\;,\;\;\;c_4=
{\z\over 2}+5\l v^2
\;\;,\;\;\;c_5=2\l v  
\;\;\;,\nonumber\\ 
&&c_6={\l\over 3}\;\;,\;\;\;c_7={\z\over2}+\l 
v^2\;\;,\;\;\;c_8={\l\over 3}
\;\;,\;\;\;c_9=
2v(\z + 2\l v^2) \;\;\;,\nonumber\\
&&c_{10}=\z + 6\l v^2\;\;,\;\;\;c_{11}=4\l v \;\;,\;\;\;
c_{12}=\l\;\;,\;\;\;c_{13}=2\l 
v \aand c_{14}=\l\;\;\;.
\label{chiggs}
\eq

The M\o ller scattering to be contemplated will 
include the scatterings
mediated by the gauge field and the Higgs ($A_m$ 
and $H$). The scattered
electrons exhibit opposite spin polarisations 
($e_{(\pm)}^-$ and $e_{(\mp)}^-$) and the same spin 
polarisations ($e_{(\pm)}^-$ and $e_{(\pm)}^-$).
The study of electrons scattered with opposite spin
polarisations is motivated by the fact that, in 
4-dimensional space-time, a Cooper
pair bound state ($s$-wave state) {\cite{cooper}} 
is built up by a scattering
between electrons correlated in phase-space with
 opposite spins. The
interactions involved in such a process are the
electromagnetic and the
phononic ones. The former is mediated by photons, 
with a repulsive behaviour,
and the latter is mediated by the phonons, which 
is attractive. The opposite
behaviour of these interactions plays a central
 r\^ole for the BCS-superconductivity phenomena 
{\cite{bcs}} 
(weak-coupling superconductors),
since, at temperatures below the critical one 
($T_c$), the interaction mediated
by phonons (attractive) is stronger
than the electromagnetic (repulsive) interaction. 
For temperatures above $T_c$,
the superconducting phase is destroyed, which means 
that the net interaction
becomes repulsive. On the other hand, the study 
of the scatterings 
between electrons with the same polarisation is 
well-motivated 
in connection with the phenomenology of superconductors 
that 
exhibit high critical magnetic fields as well as the 
re-entrant superconductivity effect. It is suspected 
that the resistence of their superconducting phase in 
presence of very high magnetic
fields is caused by $p$-electron pairing. 

For a 3-dimensional space-time, we are now trying to
 understand, with the help
of the model proposed here, what happens if we 
consider
electrons scattered in $s$ and $p$ processes. One of the 
questions to be answered is
whether or not there is a net attractive interaction 
in $e_{(\pm)}^-$$-$$e_{(\mp)}^-$ and $e_{(\pm)}^-$$-
$$e_{(\pm)}^-$
scatterings, as mediated by 
the gauge field and the
Higgs. Another interesting point to be analised 
concerns the influence of spin
polarisations ($+$ and $-$) on the dynamical nature of 
these scattering
processes.

\section{Scattering potentials}

To compute the scattering amplitudes, it will be 
necessary to derive the
Feynman rules for propagators and interaction 
vertices involving the fermions,
the gauge field and the Higgs. From the action 
(\ref{action3}), the following
propagator and vertex Feynman rules come out :
\begin{enumerate}
\item{fermions and Higgs propagators :}
\be
\langle{\ov\j _+}{\j}_+\rangle = i{{{\sl{k}}+m}
\over{k^2-m^2}}\;\;,
\;\;\;\langle{\ov\j _-}{\j}_-\rangle = i{{{\sl{k}}-m}
\over{k^2-m^2}} \aand
\langle{H H}\rangle = {i\over 2}{1\over{k^2-M_H^2}}
 \;\;\;;\label{frmp}
\ee
\item{gauge field propagator :}
\be
\langle{A_m}{A_n}\rangle = -i \left[{1\over{(k^2-M_A^2)}}
 \left(\eta_{m n} -
{k_{m} k_{n}\over M_A^2} \right)  + {1\over M_A^2}
\left({k_{m}k_{n}\over{k^2-\x M_A^2}} \right)\right] 
\;\;\;;\label{frgp}
\ee
\item{vertex Feynman rules :}
\be
\cv_{+H+}=2iyv\;\;,\;\;\;\cv_{-H-}=-2iyv\;\;,\;\;\;
\cv^m_{+A+}=iqg\gm \aand
\cv^m_{-A-}=iqg\gm \;\;\;.\label{frv}
\ee
\end{enumerate}
It should be noticed that the convention addopted, 
$\cv_{+H+}$, means the
vertex Feynman rule for the interaction term, ${\ov\j_+}
H{\j}_+$. This
convention is addopted similarly for the other interaction
 vertices above.

The $s$-channel amplitudes for the 
$e_{(\pm)}^-$$-$$e_{(\mp)}^-$  
and $e_{(\pm)}^-$$-$$e_{(\pm)}^-$ scatterings
 by the gauge field
and Higgs, are listed below :
\begin{enumerate}
\item{scattering amplitude by $A_m$ :}
\letra
\bq
&\!\!\!\!\!\!\!\!\!\!
-i{\cm}_{\pm A \mp}={\ol u}_{\pm}(p_1)
\left[iqg \g^m_{(\pm)}\right] 
u_{\pm}(p^\prime_1) \left\{
-i{\eta_{m n}\over{k^2-M_A^2}} \right\} {\ol u}_{\mp}(p_2)
\left[iqg \g^n_{(\mp)}
\right] u_{\mp}(p^\prime_2) \;\;;\label{ampa1} \\
&\!\!\!\!\!\!\!\!\!\!
-i{\cm}_{\pm A \pm}={\ol u}_{\pm}(p_1)
\left[iqg \g^m_{(\pm)}\right] 
u_{\pm}(p^\prime_1) \left\{
-i{\eta_{m n}\over{k^2-M_A^2}} \right\} {\ol u}_{\pm}(p_2)
\left[iqg \g^n_{(\pm)}
\right] u_{\pm}(p^\prime_2) \;\;;\label{ampa2}
\eq
\antiletra

\item{scattering amplitude by $H$ :}
\letra
\bq
&\!\!\!\!\!\!\!\!\!\!
-i{\cm}_{\pm H \mp}={\ol u}_{\pm}(p_1)\left[\pm 2iyv \right] 
u_{\pm}(p^\prime_1) 
\left\{ {i\over
2}{1\over{k^2-M_H^2}} \right\} {\ol u}_{\mp}(p_2)
\left[\mp 2iyv \right]
u_{\mp}(p^\prime_2) \;\;;\label{amph1} \\
&\!\!\!\!\!\!\!\!\!\!
-i{\cm}_{\pm H \pm}={\ol u}_{\pm}(p_1)\left[\pm 2iyv \right] 
u_{\pm}(p^\prime_1) 
\left\{ {i\over
2}{1\over{k^2-M_H^2}} \right\} {\ol u}_{\pm}(p_2)
\left[\pm 2iyv \right]
u_{\pm}(p^\prime_2) \;\;, \label{amph2}
\eq
\antiletra
\end{enumerate}
where $k^2$$=$$(p_1 - p^\prime_1)^2$ is the invariant 
squared momentum
transfer. The Dirac spinors, $u_+$ and $u_-$, are 
the positive-energy solutions
to the Dirac equations for $\j_+$ and $\j_-$ 
(see the Appendix), and 
they are normalised to :
\be
{\ol u}_+(p) u_+(p)=1 \aand {\ol u}_-(p) u_-(p)=-1 
\;\;\;. \label{norm}
\ee
As discussed in detail in the Appendix, we should stress here 
that the
wave functions $u_+$ and $u_-$ refer both to the particle 
(electron) 
with opposite spins, whereas $v_+$ and $v_-$ describe both the 
anti-particle
 (positron) with opposite spins. In our case, we are actually 
computing 
the scattering of 2 electrons with the opposite 
($e_{(\pm)}^-$ and $e_{(\mp)}^-$) and the same ($e_{(\pm)}^-$ 
and 
$e_{(\pm)}^-$)
spins.

To compute the scattering potentials for the interaction 
between electrons with
opposite spin polarisations ($e_{(\pm)}^-$ and $e_{(\mp)}^-$) 
and with the same spin polarisations 
($e_{(\pm)}^-$ and $e_{(\pm)}^-$), 
we refer to the works
of Sucher {\it et al.} {\cite{sucher}}, where the concept
 of potential in
quantum field theory and in scattering processes is discussed 
in great detail.

The calculation of scattering potentials will be performed in 
the
center-of-mass frame, for in this frame the electrons 
scattered are correlated
in momentum space.

By using the Feynman rules displayed above (eqs.(\ref{frmp}),
 (\ref{frgp}) and
(\ref{frv})), the following scattering potentials for the
$e_{(\pm)}^-$$-$$e_{(\mp)}^-$  
and $e_{(\pm)}^-$$-$$e_{(\pm)}^-$ scattering 
processes ($s$ and $p$-wave processes) mediated by 
the gauge field and
the Higgs are found in the center-of-mass frame ({\it{c.m.}}):
\footnote{In the
{\it c.m.} frame, the squared momentum transfer is given 
by
$k^2$$=$$-\vec{q}^{\,2}$. The notations, 
${\cu_{\pm A \mp}}(\vec r)$, ${\cu_{\pm H \mp}}(\vec r)$, 
${\cu_{\pm A \pm}}(\vec r)$ and ${\cu_{\pm H \pm}}(\vec r)$, 
with $r$$\equiv$$|\vec{r}|$, refer to 
the scattering
potentials (in configuration space) for the processes 
$e_{(\pm)}^-$$-$$e_{(\mp)}^-$ and 
$e_{(\pm)}^-$$-$$e_{(\pm)}^-$,
mediated by gauge field and Higgs. The product 
$\b_{(\pm)}\b_{(\mp)}$ is a
spinorial factor in the space of the electrons $e_{(\pm)}^-$
 and $e_{(\mp)}^-$:
$\b_{(+)}$$=$$\g_{(+)}^0$, $\b_{(-)}$$=$$-\g_{(-)}^0$ 
and
$\vec\a_{(\pm)}$$\equiv$$\g_{(\pm)}^0 \vec\g_{(\pm)}$.  }
\begin{enumerate}
\item{gauge field scattering potential :}
\letra
\bq
{\cu^{c.m.}_{\pm A \mp}}(\vec r)\!\!\!\!&=&\!\!\!\!
q^2g^2 \b_{(\pm)} 
\b_{(\mp)}
\g^m_{(\pm)} \g_m^{(\mp)} \int{d^2\vec{q}\over (2\p)^2}
{1\over{\vec{q}^{\,2} +
M_A^2}} e^{i\vec{q}.\vec{r} } \nonumber \\
\nonumber \\
&=&\!\!\!- q^2g^2 \g_{(\pm)}^0 \g_{(\mp)}^0 
\g^m_{(\pm)} \g_m^{(\mp)} 
K_0 (M_A r)
\nonumber \\
\nonumber \\
&=&\!\!\!- q^2g^2 \left[ \I - \vec\a_{(\pm)}.\vec\a_{(\mp)} 
\right] K_0 (M_A r)
 \;\;\;; \label{potA1} \\
{\cu^{c.m.}_{\pm A \pm}}(\vec r)\!\!\!\!&=&\!\!\!\!
q^2g^2 \b_{(\pm)} 
\b_{(\pm)}
\g^m_{(\pm)} \g_m^{(\pm)} \int{d^2\vec{q}\over (2\p)^2}
{1\over{\vec{q}^{\,2} +
M_A^2}} e^{i\vec{q}.\vec{r} } \nonumber \\
\nonumber \\
&=&\!\!\! q^2g^2 \g_{(\pm)}^0 \g_{(\pm)}^0 
\g^m_{(\pm)} \g_m^{(\pm)} 
K_0 (M_A r)
\nonumber \\
\nonumber \\
&=&\!\!\! q^2g^2 \left[ \I - \vec\a_{(\pm)}.\vec\a_{(\pm)} 
\right] K_0 (M_A r)
 \;\;\;; \label{potA2}
\eq
\antiletra
The minus sign in (\ref{potA1}) deserves some attention. 
It is due to the fact that
$\b_{(-)}$$=$$-\g_{(-)}^0$. This is a peculiarity of 
($1$$+$$2$)-dimensions: $\j_+$
and $\j_-$ have mass terms with opposite signs 
(therefore, opposite spins, according to
{\cite{djt,binegar}}) and so, by looking at the 
Hamiltonians displayed in the Appendix, one reads off
$\b$-terms with opposite signs.

\item{Higgs scattering potential :}
\letra
\bq
{\cu^{c.m.}_{\pm H \mp}}(\vec r)\!\!\!\!&=&\!\!\!\!2y^2v^2 
\b_{(\pm)} \b_{(\mp)}
\int{d^2\vec{q}\over (2\p)^2}{1\over{\vec{q}^{\,2} + 
M_H^2}}
e^{i\vec{q}.\vec{r} }
\nonumber \\
\nonumber \\
&=&\!\!\!-2y^2v^2 \left[ \g_{(\pm)}^0 \g_{(\mp)}^0 \right]
 K_0(M_H r)\;\;\;, \label{potH1} \\
{\cu^{c.m.}_{\pm H \pm}}(\vec r)\!\!\!\!&=&\!\!\!\!-2y^2v^2 
\b_{(\pm)} \b_{(\pm)}
\int{d^2\vec{q}\over (2\p)^2}{1\over{\vec{q}^{\,2} + 
M_H^2}}
e^{i\vec{q}.\vec{r} }
\nonumber \\
\nonumber \\
&=&\!\!\!-2y^2v^2 \left[ \g_{(\pm)}^0 \g_{(\pm)}^0 \right]
 K_0(M_H r)\;\;\;, \label{potH2}
\eq
\antiletra
\end{enumerate}
where $K_0(M r)$ is the zeroth-order modified Bessel
 function of the second
kind :
\be
 \int{d^2\vec{q}\over (2\p)^2}
{1\over{\vec{q}^{\,2} + M^2}} e^{i\vec{q}.\vec{r}
}={1\over 2\p}K_0(M r) \;\;\;. \label{k0}
\ee

This Bessel function presents the following 
asymptotic behaviour in terms of
the Compton wave-length (${1\over M}$) :
\bq
K_0(M r) \longrightarrow
&\left\{\begin{array}{l}
-\ln(M r) \;\;,\;\;\; M r\ll 1 \\
\\
\sqrt{{\p\over 2Mr}} e^{-Mr}  \;\;,\;\;\; M r\gg 1 
\;\;\;\;\;. \label{ak0}
\end{array}\right.
\eq

Now, some conditions on the parameters must be set in order to 
guarantee a
net attractive interaction between scattered electrons with opposite 
and equal spin 
polarisations, $s$ and $p$-wave scattering, respectively. 
To do that, one assumes 
the following fine tunning among the parameters :
\be
Q^2g^2 = \z + 2\l v^2 \aand q^2g^2 < 2y^2 v^2 \;\;\;.
\ee
From the conditions above, and the conditions given by 
eqs.(\ref{cond}), (\ref{vac}), (\ref{masses}) and (\ref{lowbound}), after  
some algebraic manipulations, an interesting inequality arises :
\be
{Q^2\over q^2}>{\l\over 3y^2} \;\;\;;
\ee
where it does not depend only on the fundamental constant, $g$ (the 
electromagnetic
coupling constant), but on the matter self-couplings.

The coherence length of a Cooper pair, as Cooper 
found out for the 2-electron
bound state {\cite{cooper}}, is much bigger than 
the electron Compton
wave-length ($m r\gg 1$), namely, the former is of order 
$10^4${\AA} 
and the latter of
$10^{-2}${\AA}. Therefore, for the sake of studying the
possible existence of a possible existence of $s$ and $p$ electron-pair 
condensates in the parity-preserving 
QED$_{3}$ discussed throughout this work, the strength of the net 
scattering
potentials in $s$ and $p$ scattering processes between electrons 
candidates
to built up $s$ and $p$ Cooper pairs, read
\bq
{\cv}_{c.m.}^{\,s}(r)\!\!\!\!&=&\!\!\!\!-\left[2y^2v^2 + q^2g^2\right]
\sqrt{{\p\over 2M_H r}} e^{-M_H r} \;\;\;; \label{netpot1} \\
\nonumber \\
{\cv}_{c.m.}^{\,p}(r)\!\!\!\!&=&\!\!\!\!-\left[2y^2v^2 - q^2g^2\right]
\sqrt{{\p\over 2M_H r}} e^{-M_H r} \;\;\;, \label{netpot2}
\eq
where the asymptotic approximation, $M_H r\gg 1$, is compatible 
with the
dimensions through which Cooper pair exists.

Therefore, this result shows that the net attractive
$e_{(\pm)}^-$$-$$e_{(\mp)}^-$ and $e_{(\pm)}^-$$-$$e_{(\pm)}^-$-
scattering potentials 
(\ref{netpot1}) and
(\ref{netpot2}) are non-confining, contrary to 
what happens for 
massive electrons scattered by massless gauge field, where 
the potential is 
completely confining {\cite{maris}}. 
Therefore, the
interactions mediated by the gauge field and the Higgs 
are attractive in a
scattering between electrons with opposite spin
 polarisations
($e_{(\pm)}^-$$-$$e_{(\mp)}^-$-scattering). For 
scatterings with a scalar exchange,
the spin polarisations do not affect the behaviour 
of potential: it will be
always attractive. This result is expected, since 
the Higgs particle does not
{\it feel} the electron polarisations.

An interesting point to remark is that, in spite the 
scattered particles have
the same electric charge, the spin polarisation is 
determinant for the
behaviour of the scattering potential for processes 
where a gauge field is
exchanged. In the case where the scattered electrons 
have opposite spin polarisations 
(the opposite mass term in Dirac's equation)
, the interaction is attractive. However, one should notice 
that this
result is not conflicting with QED expectations. In our model, 
the photon-mediated interaction takes place on a non-trivial
background, set by the Higgs field. Electron interaction is 
repulsive
if the exchanged photon propagates on a QED vacuum. In the
 model proposed
here, the electron mass and the electron interactions are 
to be referred
not to a trivial vacuum, but to a background responsible 
for the photon 
mass. Therefore, we interpret the attraction as a 
byproduct of the 
physics of electrons propagating on a non-trivial 
Higgs background. 
Nevertheless, for scatterings
between electrons with the same spin state 
(the same mass term in Dirac's equation), the 
interaction becomes repulsive.

\section{Discussions and general conclusions}

In this work we concentrate efforts in trying to 
understand 
many intriguing features of dynamical processes in {\Ddd}, 
where a parity-preserving QED$_3$ is coupled to scalar 
matter. In this scenario, the spontaneous symmetry breaking 
mechanism of a $U(1)$-symmetry takes place. The breakingdown 
is realised by a sixth-power potential with a mass 
generation for the gauge boson and the fermions. The 
spontaneous symmetry breaking mechanism is the responsible
for the appearance of a kind of Meissner effect, since 
the scalar magnetic 
field obeys a London equation: $B(r)$$=$$B_0 e^{-\l r}$, 
where 
$\l$$=$${1\over M_A}$ is the penetration length. Therefore,
as $M_A$ depends on the vacuum expectation value, $v$, it 
can be concluded that $\l$$\rightarrow$$\infty$ when 
$v$$\rightarrow$$0$. This means that the Meissner effect 
is completely destroyed when the scalar assumes a vanishing 
vacuum expectation value, as in the symmetric regime. 

Now,
bering in mind the Coleman-Mermin-Wagner theorem {\cite{col}}, 
an interesting
question naturally comes out: is there a critical temperature, 
$T_c$, such that for temperatures above the critical one 
($T>T_c$) the gauge symmetry is restored? 
If yes, it follows that
the scalar assumes a vanishing vacuum expectation value which
leaves the gauge field massless and, as a consequence, the 
Meissner effect disappears. Therefore, a 
superconducting-type phase transition should be present as 
a direct consequence of symmetry restoration by the 
finite temperature quantum corrections. 

An interesting point to be emphasised is the influence 
of spin polarisations on the dynamical nature of the scattering
processes. This feature is dictated by the Poincar\'e group 
structure of {\Ddd}. As a peculiarity of this space-time
and the Higgs background on which the electrons and photons
propagate, 
electrons scattered by a massive 
gauge boson and by a Higgs can experience an attractive 
interaction. The interaction potential associated to gauge
boson exchange displays opposite behaviours when the electrons
scattered have opposite or the same spin polarisations, since 
electrons
propagate on a non-trivial Higgs background. For 
the case where the Higgs is exchanged, the scattering 
potential is completely insensitive to the electrons
polarisation, as is expected, since the Higgs is a
spinless particle. One concludes in this work that 
electrons can attract each other in {\Ddd} through 
scattering processes where a massive gauge boson and a 
Higgs are involved. This attraction between electrons might 
favour a bound state. As long as 
the behaviour of this model at the quantum level is 
concerned, it 
shows to be stable under radiative corrections and anomaly
free in the symmetric and broken regimes, 
which proves its renormalisability {\cite{arqed3}}.

It should be pointed out that, in order to be sure of 
the existence of a bound
state in such scatterings, it is more advisable to 
study the Bethe-Salpeter
{\cite{bethesalp}} equation in {\Ddd} {\cite{bethesalp3}}
 for the model
proposed here. Such an analysis is more reliable in view 
of its intrinsically
non-perturbative nature. It is worthwhile to stress that 
our results simply
suggest that, at the semiclassical level, a net 
attractive interaction between
electrons with opposite polarisations might point 
out pair condensation if
Bethe-Salpeter equations are taken into account 
{\cite{bethesalp3}}. On the other hand, if an attraction 
is felt at the level of tree amplitudes, 
we would not expect
that loop corrections, that bring about powers of 
$\hbar$, might work against
pair condensation. In any case, to our mind, it
 would be more reasonable to
pursue an investigation of the Bethe-Salpeter equations 
(rather than computing
higher-loop corrections) in order to infer about
 electron-pair condensation in
the model discussed throughout this paper {\cite{bsqed3ssb}}. 

As a final remark, we point 
out that the finite temperature approach 
could be of interest in order 
to verify whether or not there are 
pair-condensation phase 
transitions for some critical temperature, $T_c$, 
in the cases of $s$ and $p$ bound states. 
If no more bound states exist 
in the solutions of the Bethe-Salpeter equation for 
temperatures above the critical one ($T>T_c$), the electrons 
are no more  
correlated and, therefore, the gauge symmetry 
is restored; as a consequence, the Meissner 
effect disappears. 

\appendix
\section{Some properties of Dirac spinors in $D$$=$$3$}
\setcounter{equation}{0}
In this Appendix, we present some aspects of Dirac spinors 
living in $D$$=$$3$, like the positive and negative energy
solutions 
to the Dirac equations satisfied by $\j_+$ and $\j_-$.  
We state clearly the connection between mass and spin and, 
in order to elucidate some peculiarities of electrons
scattering in 3 space-time dimensions, we present the 
Hamiltonian for both $\j_+$ and $\j_-$. We also compute 
explicitly the
charges of the positive and negative energy wave functions 
associated to 
$\j_+$ and $\j_-$

\subsection{Positive and negative energy solutions 
for $\j_+$ and $\j_-$}
Let us consider $u_+$ and $v_+$, $u_-$ and $v_-$, 
respectively, as the positive 
and negative solutions to the Dirac equations for 
$\j_+$ and $\j_-$. Therefore, they satisfy the 
following equations in momentum space :
\bq
&&({\sl{p}} - m) u_+(p) = 0 \;\;,\;\;\;
(-{\sl{p}} - m) v_+(p) = 0 \;\;\;; \label{Diraceq+} \\ 
&&({\sl{p}} + m) u_-(p) = 0 \;\;,\;\;\;
(-{\sl{p}} + m) v_-(p) = 0 \;\;\;. \label{Diraceq-}
\eq
Their solutions are given by
\bq
&&u_+(p)={ {{\sl{p}} + m}\over{\sqrt{2m(m+E)}} }
\;u_+(m,\vec{0})\;\;,\;\;\;
v_+(p)={ -{{\sl{p}} + m}\over{\sqrt{2m(m+E)}} }
\;v_+(m,\vec{0})\;\;\;; \label{gensolu1}\\
&&u_-(p)={ -{{\sl{p}} + m}\over{\sqrt{2m(m+E)}} }
\;u_-(m,\vec{0})\;\;,\;\;\;
v_-(p)={ {{\sl{p}} + m}\over{\sqrt{2m(m+E)}} }
\;v_-(m,\vec{0})\;\;\;, \label{gensolu2}
\eq
where $E\equiv k^0={\sqrt{ {\vec{k}}^2 + m^2 }}>0$. 
The wave functions 
$u_+(m,\vec{0})$, $v_+(m,\vec{0})$, $u_-(m,\vec{0})$ and 
$v_-(m,\vec{0})$ are the solutions of 
eqs.(\ref{Diraceq+}-\ref{Diraceq-}) in the rest frame 
\bq
&&u_+(m,\vec{0})={1\over\sqrt{2}}
\left(\begin{array}{c}
1\\
1
\end{array}\right) \;\;,\;\;\;
v_+(m,\vec{0})={1\over\sqrt{2}}
\left(\begin{array}{c}
1\\
-1
\end{array}\right) \;\;\;;\label{restsolu+}\\
&&u_-(m,\vec{0})={1\over\sqrt{2}}
\left(\begin{array}{c}
1\\
-1
\end{array}\right) \;\;,\;\;\;
v_-(m,\vec{0})={1\over\sqrt{2}}
\left(\begin{array}{c}
1\\
1
\end{array}\right) \;\;\;.\label{restsolu-}
\eq

The positive and negative energy solutions given by 
eqs.(\ref{gensolu1}-\ref{gensolu2}) are normalised to :
\bq
&&{\ol u}_+(p) u_+(p)=1\;\;,\;\;\;
{\ol v}_+(p) v_+(p)=-1 \;\;\;;\label{norm1}\\ 
&&{\ol u}_-(p) u_-(p)=-1\;\;,\;\;\;
{\ol v}_-(p) v_-(p)=1 \;\;\;. \label{norm2}
\eq

\subsection{The spin of $u_+$, $v_+$, 
$u_-$ and $v_-$ }
Now, by considering the results of last subsection, one is able
to determine the spins of the solutions 
$u_+$, $v_+$, $u_-$ and $v_-$. We compute the spins in the 
particle rest frame, since we have in mind to explicitly 
exhibit the fact that the sign of the mass term fixes the 
polarisation of the fermion.

In $D$$=$$3$, the generators of the $\ov{SO(1,2)}$ group in 
the spinor representation read :
\be
\Sigma^{kl}={\ts\frac14}\;[\gamma^k,\gamma^l]\;\;\;,\label{skl}
\ee
where the $\g$-matrices are taken as
$\g^m$$=$$(\sx,i\sy,-i\sz)$. 

The spin operator $S^{12}$ is obtained from (\ref{skl}), and 
it reads
\be
S^{12}={1\over 2}\;\sx\;\;\;.\label{spin}
\ee
Its action upon the rest frame wave functions 
given by eqs.(\ref{restsolu+}-\ref{restsolu-}) is collected 
below :
\bq
&&S^{12}\;u_+(m,\vec{0})=s^u_+ \;u_+(m,\vec{0}) \;\;,\;\;\;
S^{12}\;v_+(m,\vec{0})=s^v_+ \;v_+(m,\vec{0})\;\;\;;\label{eigens1} \\
&&S^{12}\;u_-(m,\vec{0})=s^u_- \;u_-(m,\vec{0})\;\;,\;\;\;
S^{12}\;v_-(m,\vec{0})=s^v_- \;v_-(m,\vec{0})\;\;\;.
\label{eigens2}
\eq

With the help of (\ref{restsolu+}-\ref{restsolu-}) and
(\ref{spin}), we find the following values for the spin eigenvalues 
$s^u_+$, $s^u_-$, $s^v_+$, and $s^v_-$ : 
\be
s^u_+={1\over 2}\;\;,\;\;\;s^u_-=-{1\over 2}\;\;,\;\;\;
s^v_+=-{1\over 2}\;\;,\;\;\;s^v_-={1\over 2}
\;\;\;.\label{spins}
\ee

From eq.(\ref{spins}), it can be concluded that electrons
($u_+$ and $u_-$) and
positrons ($v_+$ and $v_-$) with opposite mass terms have 
opposite spin 
polarisations. It should be pointed out that this result is
in completely agreement with ref.{\cite{binegar}}. In the 
Section A.5 of this Appendix we prove explicitly that the 
wave functions $u_+$ and $u_-$ are associated to electrons 
whereas $v_+$ and $v_-$ are associated to positrons. 

An interesting point to stress here concerns the 
polarisations
of a particle ($u$) and the corresponding anti-particle 
($v$)
belonging to the same Dirac spinor ($\j$). As a typical
feature of 3 space-time dimensions, if a particle has 
spin $s$, 
its anti-particle has spin $-s$.  

\subsection{The Hamiltonian for $\j_+$ and $\j_-$}

In this subsection, the relation between the 
opposite mass term signs and 
the opposite signs of $\b$-matrices respected to $\j_+$ 
and $\j_-$ 
becomes clear by computing the free Hamiltonian 
operator $H_0$.

For a general massive Dirac spinor, $\c$, the free 
Hamiltonian operator in momentum space,
$H_0$, is given by :
\be
H_0 \c{\equiv}({\vec \a}.{\vec p} + \b m) \c \;\;\;,\label{H0}
\ee
where  
\be
{\vec \a}=\g^0 {\vec \g} \aand \b=\g^0 \;\;\;.
\ee

Now, considering the Dirac equations for $\j_+$ and $\j_-$ :
\be
(i{\sl{\pa} - m}) {\j}_+ =0 \aand 
(i{\sl{\pa} + m}) {\j}_- =0 \;\;\;,
\ee
it follows that
\bq
&&i{\pa\over{\pa t}}{\j}_+=\left(i\g_{(+)}^0\vec\g_{(+)}.
\vec{\pa} 
+ \b_{(+)} m\right){\j}_+ \equiv H_0^{(+)}{\j}_+ \;\;\;;\\
&&i{\pa\over{\pa t}}{\j}_-=\left(i\g_{(-)}^0\vec\g_{(-)}.
\vec{\pa}
+ \b_{(-)} m\right){\j}_- \equiv H_0^{(-)}{\j}_-\;\;\;.
\eq
Therefore, in momentum space, the Hamiltonians $H_0^{(+)}$ and 
$H_0^{(-)}$ read
\bq
&&H_0^{(+)}{\j}_+=\left({\vec \a_{(+)}}.{\vec p} + \b_{(+)} m 
\right){\j}_+ \;\;\;; \label{H+}\\ 
&&H_0^{(-)}{\j}_-=\left({\vec \a_{(-)}}.{\vec p} + \b_{(-)} m 
\right){\j}_- \;\;\;,\label{H-} 
\eq
where, from (\ref{H0}), it can be concluded that
\bq
&&{\vec \a_{(+)}}=\g_{(+)}^0\vec\g_{(+)} 
\aand \b_{(+)}=\g_{(+)}^0 \;\;\;; \label{b1}\\
&&{\vec \a_{(-)}}=\g_{(-)}^0\vec\g_{(-)}
\aand \b_{(-)}=-\g_{(-)}^0 \;\;\;. \label{b2}
\eq

The eqs.(\ref{b1}-\ref{b2}) completely determine the scattering 
potentials behaviour for the scattering processes
of $e_{(\pm)}^-$$-$$e_{(\mp)}^-$  
and $e_{(\pm)}^-$$-$$e_{(\pm)}^-$ mediated by the gauge field and 
the Higgs. They are in agreement to the fact that the 
Higgs scattering potential does not {\it feel} 
the electron polarisations, since Higgs is spinless. It is 
only possible if eqs.(\ref{b1}-\ref{b2}) are fulfiled. 

\subsection{The spin of $u_+$, $v_+$, 
$u_-$ and $v_-$ as a quantum number}

Let us consider the spin operator given by eq.(\ref{spin}) :
\begin{eqnarray*} 
&&S^{12}={1\over 2}\;\sx\;\;\;,
\end{eqnarray*} 
and the free Hamiltonian 
operators in momentum space for the spinors ${\j}_+$ and ${\j}_-$ 
(eqs.(\ref{H+}-\ref{H-})) : 
\bq
&&H_0^{(+)}=\left({\vec \a_{(+)}}.{\vec p} + \b_{(+)} m 
\right)\;\;\;,\nonumber \\ 
&&H_0^{(-)}=\left({\vec \a_{(-)}}.{\vec p} + \b_{(-)} m 
\right) \;\;\;,\nonumber 
\eq 
where ${\vec \a_{(\pm)}}$ and $\b_{(\pm)}$ are given 
by eqs.(\ref{b1}-\ref{b2}). It can be easily shown that the 
following commutators vanish
\bq
&&\left[H_0^{(+)},S^{12}\right]=0 \;\;\;, \\
&&\left[H_0^{(-)},S^{12}\right]=0 \;\;\;.
\eq
This result ensures that the eigenvalues ($s^u_+$, $s^v_+$, 
$s^u_-$ and $s^v_-$) of the spin operator, $S^{12}$, 
corresponding respectively to the wave functions $u_+$, $v_+$, 
$u_-$ and $v_-$ are indeed good quantum numbers to label 
physical states. 

\subsection{The charges of $u_+$, $v_+$, 
$u_-$ and $v_-$}

In order to determine the charges of the particles 
associated to the wave functions, $u_+$, $v_+$, $u_-$ 
and $v_-$, it is necessary to compute the eigenvalues 
of the charge operators, $Q_+$ and $Q_-$, respected to 
the field operators, ${\j}_+$ and ${\j}_-$. Their 
expansion in terms of the creation and annihilation operators 
read as below :
\bq
&&{\j}_+(x) =  \int{d^2\vec{k}\over (2\p)^2}{m\over k^0}
\left[a_+(k) u_+(k) e^{-ik.x} + b_+^{\dg}(k) v_+(k) e^{ik.x} 
\right] \;\;\;,\label{exp1}\\
&&{\j}_-(x) = \int{d^2\vec{k}\over (2\p)^2}{m\over k^0}
\left[a_-(k) u_-(k) e^{-ik.x} + b_-^{\dg}(k) v_-(k) e^{ik.x} 
\right]\;\;\;,\label{exp2}\\
&&{\ol\j}_+(x) = \int{d^2\vec{k}\over (2\p)^2}{m\over k^0}
\left[a_+^{\dg}(k) {\ol u}_+(k) e^{ik.x} + b_+(k) {\ol v}_+(k) 
e^{-ik.x} 
\right]\;\;\;,\label{exp3}\\
&&{\ol\j}_-(x) = \int{d^2\vec{k}\over (2\p)^2}{m\over k^0}
\left[a_-^{\dg}(k) {\ol u}_-(k) e^{ik.x} + b_-(k) {\ol v}_-(k) 
e^{-ik.x} \right]\;\;\;,\label{exp4}
\eq
where the operators, $a_+^{\dg}$, $b_+^{\dg}$, $a_-^{\dg}$ and 
$b_-^{\dg}$, are the creation operators, and, $a_+$, $b_+$, 
$a_-$ and $b_-$, are the annihilation operators. The wave 
functions were analysed in details in the Section A.1 of 
this Appendix. 

With the help of the Dirac equations (\ref{Diraceq+}-
\ref{Diraceq-}), the normalisation conditions (\ref{norm1}-
\ref{norm2}) and the relation 
\be
\left\{ {\sl p},\g^0 \right\}=2p^0 \;\;\;,
\ee
the following equations are satisfied by the wave functions 
$u_+$, $v_+$, $u_-$ and $v_-$ :
\bq
&&{u}_+^{\dg}(p) u_+(p)={p^0\over m}\;\;,\;\;\;
{v}_+^{\dg}(p) v_+(p)={p^0\over m}\;\;\;;\label{norm1+}\\ 
&&{u}_-^{\dg}(p) u_-(p)={p^0\over m}\;\;,\;\;\;
{v}_-^{\dg}(p) v_-(p)={p^0\over m}\;\;\;. \label{norm2-}
\eq

The microcausality fixes the following anticommutation relations :
\be
\left\{{\j}_+(x),{\j}_+^{\dg}(y) \right\}_{x^0=y^0}=
\d^2(\vec{x}-\vec{y})\;\;,\;\;\;
\left\{{\j}_-(x),{\j}_-^{\dg}(y) \right\}_{x^0=y^0}=
\d^2(\vec{x}-\vec{y})\;\;\;.\label{micro}
\ee
Now, by assuming the field operator expansions 
(\ref{exp1}-\ref{exp4}), and the normalisation conditions 
given by eqs.(\ref{norm1+}-\ref{norm2-}), the anticommutation 
relations between the creation and annihilation operators read :
\bq
&&\left\{a_+(k),a_+^{\dg}(p) \right\}= (2\p)^2~{k^0\over m}
~\d^2(\vec{k}-\vec{p})\;\;\;, \label{a+a+}\\
&&\left\{b_+(k),b_+^{\dg}(p) \right\}= (2\p)^2~{k^0\over m}
~\d^2(\vec{k}-\vec{p})
\;\;\;,\label{b+b+} \\
&&\left\{a_-(k),a_-^{\dg}(p) \right\}= (2\p)^2~{k^0\over m}
~\d^2(\vec{k}-\vec{p})\;\;\;,\label{a-a-}\\
&&\left\{b_-(k),b_-^{\dg}(p) \right\}= (2\p)^2~{k^0\over m}
~\d^2(\vec{k}-\vec{p})
\;\;\;.\label{b-b-}
\eq

The charge operators, $Q_+$ and $Q_-$, associated to the field 
operators, ${\j}_+$ and ${\j}_-$, are defined by the following 
normal ordering products :
\bq
&&Q_+= \int{d^2\vec{x}}:j_+^o(x):=-qg \int{d^2\vec{x}}:
{\j}_+^{\dg}(x){\j}_+(x): 
\;\;\;,\label{j+}\\
&&Q_-= \int{d^2\vec{x}}:j_-^o(x):=-qg \int{d^2\vec{x}}:
{\j}_-^{\dg}(x){\j}_-(x): 
\;\;\;,\label{j-}
\eq
which in terms of the creation and annihilation operators 
are given by
\bq
&&Q_+=-qg \int{d^2\vec{k}\over (2\p)^2}{m\over k^0}
\left[a_+^{\dg}(k) a_+(k) -
b_+^{\dg}(k) b_+(k) \right] \;\;\;, \label{Q+}\\
&&Q_-=-qg \int{d^2\vec{k}\over (2\p)^2}{m\over k^0}
\left[a_-^{\dg}(k) a_-(k) -
b_-^{\dg}(k) b_-(k) \right] \;\;\;. \label{Q-}
\eq

From the anticommutation relations (\ref{a+a+}-\ref{b-b-}) 
and the eqs.(\ref{Q+}-\ref{Q-}), for the charge operators 
$Q_+$ and $Q_-$, it can be easily shown that
\bq
&&\left[Q_+,a_+^{\dg}(p) \right]=-qg~a_+^{\dg}(p)\;\;,\;\;\;
\left[Q_+,b_+^{\dg}(p) \right]=+qg~b_+^{\dg}(p)\;\;\;, 
\label{comm+}\\
&&\left[Q_-,a_-^{\dg}(p) \right]=-qg~a_-^{\dg}(p)\;\;,\;\;\;
\left[Q_-,b_-^{\dg}(p) \right]=+qg~b_-^{\dg}(p)\;\;\;.
\label{comm-}
\eq

Let us denote the vacuum ground state by the ``ket'', $|0\ra$, 
such that
\bq
&&a_+(k)|0\ra=0 \;\;,\;\;\; b_+(k)|0\ra=0 \;\;\;, \\
&&a_-(k)|0\ra=0 \;\;,\;\;\; b_-(k)|0\ra=0 \;\;\;,
\eq
where $\la0|0\ra=1$. Now, bearing in mind the commutation 
relations given by eqs.(\ref{comm+}-\ref{comm-}), and applying 
them to the vacuum state, it follows that
\bq
&&Q_+|e_{(+)}^-\ra=-qg~|e_{(+)}^-\ra 
~~~\mbox{where}~~~|e_{(+)}^-\ra = a_+^{\dg}|0\ra \;\;\;;\\
&&Q_+|e_{(+)}^+\ra=+qg~|e_{(+)}^+\ra 
~~~\mbox{where}~~~|e_{(+)}^+\ra = b_+^{\dg}|0\ra \;\;\;; \\
&&Q_-|e_{(-)}^-\ra=-qg~|e_{(-)}^-\ra 
~~~\mbox{where}~~~|e_{(-)}^-\ra = a_-^{\dg}|0\ra \;\;\;;\\
&&Q_-|e_{(-)}^+\ra=+qg~|e_{(-)}^+\ra 
~~~\mbox{where}~~~|e_{(-)}^+\ra = b_-^{\dg}|0\ra \;\;\;.
\eq
Due to these results, one concludes that :
\begin{enumerate}
\item{$a_+^{\dg}$ creates an electron ($u_+$) with spin 
$s^u_+={1\over 2}$ and charge $-qg$.}
\item{$b_+^{\dg}$ creates a positron ($v_+$) with spin 
$s^v_+=-{1\over 2}$ and charge $+qg$.}
\item{$a_-^{\dg}$ creates an electron ($u_-$) with spin 
$s^u_-=-{1\over 2}$ and charge $-qg$.}
\item{$b_-^{\dg}$ creates a positron ($v_-$) with spin 
$s^v_-={1\over 2}$ and charge $+qg$.}
\end{enumerate}

As a final conclusion, $u_+$ and $u_-$ are wave functions of 
electrons with opposite spins 
($e_{(+)}^-$ and $e_{(-)}^-$), whereas $v_+$ and $v_-$ are 
wave functions of positrons with 
opposite spins ($e_{(+)}^+$ and $e_{(-)}^+$), which is in 
completely agreement with the fact that spin
is related to a space-time symmetry (Lorentz group) and electric 
charge is related to an
internal symmetry (gauge symmetry). Some of the physical relevant 
results obtained in this Appendix are summarised in 
Table~{\ref{table1}}.
\begin{table}[hbt]
\centering
\begin{tabular}{|c|c|c|c|c|c|c|}
\hline
   Creation  &Charge  &Charge &Particle &Symbol &Wave &Spin  \\ 
   operator &operator & & & &function & \\
\hline\hline
$a_+^{\dg}$ &$Q_+$ &$-qg$ &electron &$e_{(+)}^-$ &$u_+$ 
&$s^u_+=+{1\over 2}$    \\
\hline
$a_-^{\dg}$ &$Q_-$ &$-qg$ &electron &$e_{(-)}^-$ &$u_-$ 
&$s^u_-=-{1\over 2}$   \\
\hline
$b_+^{\dg}$ &$Q_+$ &$+qg$ &positron &$e_{(+)}^+$ &$v_+$ 
&$s^v_+=-{1\over 2}$   \\
\hline
$b_-^{\dg}$ &$Q_-$ &$+qg$ &positron &$e_{(-)}^+$ &$v_-$ 
&$s^v_-=+{1\over 2}$   \\
\hline
\end{tabular}
\caption[t1]{Charge and spin of the particles associated to the field 
operators, ${\j}_+$ and ${\j}_-$.}
\label{table1}
\end{table}

All results presented in this Appendix show non-trivial aspects of 
parity-preserving QED$_3$. The relation between the signal of spin and 
the signal of mass in the Dirac mass term is an interesting feature of 
massive fermions in {\Ddd}. Another point is the fact that, due to Higgs 
mechanism, the interaction potential experienced by the electrons of both 
possible polarisations are non-confining, contrary to the case where 
massless gauge field are taken into account {\cite{maris}}. It provides 
a net attractive interaction between electrons of both spins, which 
might favour an electron-pair condensation of $s$ and $p$-wave type. 

\small

\subsection*{Acknowledgements}
The authors express their gratitude to Prof. O. Piguet, Dr. A.P. Demichev, 
Dr. D.H.T. Franco, Prof. K.S. Narain, Prof. V.L. Baltar and Dr. E. Mucciolo 
for nice remarks and
helpful discussions, and to Prof. G.W. Semenoff and Prof. J. Sucher for their
 comments and
suggestions. One of the authors (O.M.D.C.) dedicates 
this work to his wife,
Zilda Cristina, and to his daughter, Vittoria, who was 
born in 20 March 1996. He also thanks to the Organizing Committee of 
{\it Quantum Systems: New Trends and Methods 96 - QS96 - Minsk - Belarus} 
and the {\it High Energy Section} of the {\it ICTP - Trieste - Italy}, 
where part of this work was done, for the kind hospitality and financial 
support, and to its Head, Prof. S. Randjbar-Daemi. Thanks are also 
due to the Head of CFC-CBPF,
Prof. A.O. Caride, for encouragement.  CNPq-Brazil is 
acknowledged for
invaluable
financial help.

\end{document}